\documentclass[aps,pre,twocolumn,superscriptaddress]{revtex4-2}
\usepackage[dvips]{graphicx}
\usepackage{amsmath,amssymb}

\def\vector#1{\mbox{\boldmath $#1$}}

\begin{document}
\title{Phase diagram of the dipolar Ising ferromagnet on a kagome lattice}

\author{Hisato \surname{Komatsu}}
\email[Email address: ]{komatsu.hisato@nims.go.jp}

\affiliation{Research Center for Advanced Measurement and Characterization, National Institute for Materials Science, Tsukuba, Ibaraki 305-0047, Japan}

\author{Yoshihiko Nonomura}
\email[Email address: ]{nonomura.yoshihiko@nims.go.jp}

\affiliation{International Center for Materials Nanoarchitectonics,  National Institute for Materials Science,  Tsukuba, Ibaraki 305-0044, Japan}

\author{Masamichi Nishino}
\email[Email address: ]{nishino.masamichi@nims.go.jp} 

\affiliation{Research Center for Advanced Measurement and Characterization, National Institute for Materials Science, Tsukuba, Ibaraki 305-0047, Japan}

\affiliation{Elements Strategy Initiative Center for Magnetic Materials, National Institute for Materials Science, Tsukuba, Ibaraki 305-0047, Japan}

\begin{abstract}
We study the field--temperature phase diagram of the two-dimensional dipolar Ising ferromagnet on a kagome lattice with a specific ratio between the exchange and dipolar constants, $\delta = 1$. 
Using the stochastic cutoff (SCO) $O(N)$ Monte Carlo method, we calculated order parameters for stripe and bubble phases and other thermodynamical quantities. 
We find two kinds of stripe phases at low fields, where the arrangement of the branch spins neighboring the stripe frame varies, and two bubble phases at high fields, in which three-spin domains (bubbles) form a regular triangular lattice but 
the triangular array of bubbles changes on a kagome lattice. 
We also find that with increasing the field, there exist a disordered phase between the stripe and bubble phases and between the two bubble phases. 
We discuss the details of the features of these phases and phase transitions.   
\end{abstract}
\maketitle

\section{Introduction}

Magnetic films have industrial importance because of possible applications in magnetic storage devices~\cite{Heinrich,Bader}. 
In such systems the competition between short-range exchange and long-range dipolar interactions with the influence of other interactions causes rich magnetic structures. For example, with magnetocrystalline anisotropies, various stipe patterns appear and spin-reorientation (SR) transitions take place~\cite{Pappas,Allenspach,Ramchal,Won,Qiu}, and with DM interactions, helical structures and skyrmions are formed~\cite{Yu}. 

In ferromagnetic films, a characteristic phenomenon has often been observed experimentally with increasing the external magnetic field. A stripe phase at low fields changes to a phase called  ``bubble phase" before the transition to a ferromagnetic phase at high fields. In the bubble phase, magnetic domains are arranged almost periodically, forming a triangular lattice~\cite{Cape,Huber,Seul}. 

The properties of the stripe and bubble phases have also been theoretically studied by magnetostatic approaches~\cite{Cape,Huber}, coarse-grained effective free energy approaches such as Ginzburg-Landau (GL) theory~\cite{Garel,Cannas2011,Mendoza} and lattice models~\cite{Mendez,Cannas2011}. 
In the magnetostatic approaches, the magnetostatic energy was compared between the structures of a parallel-stripe array and cylindrical domains, and a field-thickness phase diagram without thermal effect was studied~\cite{Cape,Huber}. 
In the effective free energy approaches, the phase boundaries between the 
stripe and bubble phases and between the bubble and uniform (ferromagnetic) phases were investigated, and a first-order transition was suggested for the former~~\cite{Garel,Cannas2011}, while a Berezinskii-Kosterliz-Thouless (BKT)-like transition~\cite{Berezinskii,Kosterlitz} was pointed out for the latter~\cite{Garel}.  
These coarse-grained approaches can treat large systems, but they were based on the mean-field theory and the thermal fluctuation effect was insufficiently treated. In addition, periodic structures were assumed for the stripe and bubble phases and the stability between the two phases were compared. 

On the other hand, in the lattice model approaches using the two-dimensional (2D) dipolar Ising and Heisenberg ferromagnets, which treat the thermal fluctuation effect precisely using Monte Carlo (MC) methods, etc., such periodic structures are spontaneously formed without the assumption of the periodicity. 
However, it is difficult to simulate the models with large sizes. Because of the long-range nature of the dipolar interaction, $O(N^2)$ ($N$ is the total number of spins) computational time, namely a high computational cost, is required. 
Stripe phases and SR transitions have been investigated in several parameters of the 2D dipolar Ising~\cite{Booth,MacIsaac-Ising,Toloza,Rastelli06,Cannas,Pighin-Ising,Rastelli,Vindigni,Rizzi,Fonseca,Ruger,Horowitz,Bab,Komatsu1}  and Heisenberg~\cite{Pescia,Moschel,Hucht,MacIsaac1,MacIsaac2,Bell,Santamaria,Rapini,Whitehead,Carubelli,Mol,Pighin2,Pighin,Mol2,Komatsu2,Komatsu3} ferromagnets, but studies on the stripe--bubble and bubble--ferromagnetic transitions are limited~\cite{Mendez,Cannas2011}. 
In these studies on a square lattice, the intermediate phase located between (anharmonic) stripe and saturated ferromagnetic phases was named bubble phase, in which domains were observed. However, the formation of any lattice structure of the domains was not studied and is unclear. 
There a first-order transition was indicated between the stripe and bubble phases, and BKT-like melting behavior was suggested between the bubble and saturated ferromagnetic phases~\cite{Mendez}. 

In the present paper, we investigate the properties of the stripe and bubble phases 
and phase transitions in the 2D dipolar Ising ferromagnet on a kagome lattice. According to the studies by the effective energy approaches, we define an order parameter for bubble phases to detect a lattice on which magnetic domains are arranged. 
We estimate several order parameters by a MC method and study the field--temperature phase diagram with a specific ratio between the exchange and dipolar constants, $\delta = 1$. 

To overcome the difficulty of  $O(N^2)$ computational time required for conventional MC algorithms, we use the stochastic cutoff (SCO) $O(N)$ MC method~\cite{Sasaki} to reduce the computational cost.
We show two stripe phases at low fields and two bubble phases at high fields, and present a disordered phase between the stripe and bubble phases, between the two bubble phases, and above the higher-field bubble phase. 
We also discuss the phase transitions associated with these phases. 
The argument on the previously suggested BKT-like melting behavior is out of scope of the present work, because it is a delicate issue which requires huge-scale computations to obtain a convincing result. 

The rest of the paper is organized as follows. In Sec.~\ref{model}, the model and method are presented. In Sec.~\ref{results}, the results and discussion are given.  
After an overview of the field--temperature phase diagram in Sec.~\ref{phase_diag}, the features of the bubble phases and stripe phases are discussed in Sec.~\ref{bubble} and Sec.~\ref{stripe}, respectively. The properties of the phase transitions are studied in Sec.~\ref{PT}. Section \ref{summary} is devoted to the summary.

\begin{figure}[thbp]
\begin{center}
\includegraphics[width=5cm]{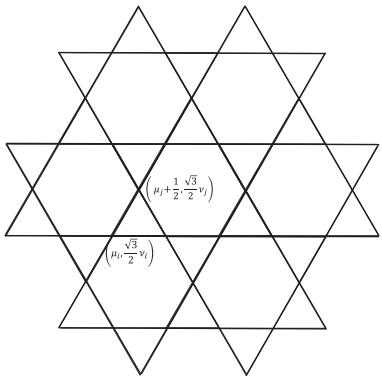}  
\end{center}
\caption{Structure of kagome lattice.}
\label{Kagome_lattice}
\end{figure}

\section{Model and method \label{model}}

We consider an Ising spin system on a kagome lattice in the $xy$ plane (Fig.~\ref{Kagome_lattice}). The position of the $i$th spin in units of the lattice constant is given as 
\begin{equation}
\left( \mu _i + \frac{1}{2} \delta_{1,\; (\nu _i \;{\rm mod}\;  2)} , \ \frac{\sqrt{3}}{2} \nu _i \right),
\label{position_spin}
\end{equation}
where $\delta_{m,n}$ is the Kronecker delta, and $\mu _i$ is integer (even) if $\nu_i$ is even (odd). 
The Hamiltonian of the system consists of the nearest-neighbor ferromagnetic Ising interaction, dipolar interaction, and Zeeman term:   
\begin{equation}
{\cal H} = - \delta \sum _{\left< i,j \right> } \sigma _i \sigma _j + \sum _{i < j  } \frac{\sigma _i \sigma _j }{r_{ij} ^3} - H \sum _i \sigma _i. 
\label{Hamiltonian}
\end{equation}
Here, $\sigma _{i}$ takes $\sigma _{i}=1$ (up spin) or $-1$ (down spin), perpendicular to the $xy$ plane, $\delta (> 0)$ is the ratio between the exchange and dipolar constants,  $r_{ij}$ is the distance between the $i$th and $j$th spins, and $H$ is the magnetic field parallel to the Ising spins. In this paper, we study the case of $\delta = 1$. 


As mentioned in the introduction, $O(N^2)$ simulation time is required in conventional MC methods, and we use the stochastic cut-off (SCO) $O(N)$ MC method~\cite{Sasaki} with a modification~\cite{Komatsu1}.  
In the original SCO algorithm, the stochastic potential switching (SPS) procedure~\cite{Mak05,Mak07} with $O(N)$ switching time is applied to all long-range interactions in the system. 
In the case of Ising spins, however, the application of the SCO method to neighboring spins causes a delay of the relaxation time, 
and a tuning of the number of dipolar interactions to which the SPS procedure is applied is necessary to accelerate the relaxation time (Appendix B in Ref.~\cite{Komatsu1}). 
In the present study, we tune the number of dipolar interactions using a Metropolis algorithm for the dipolar interactions which satisfy $|\mu_i - \mu_j| \le 5$ and $|\nu_i - \nu_j| \le 5$, and the SCO algorithm for the dipolar interactions outside this region.

To estimate the $H$ dependences of the order parameters at a given $T$, 
a simulated annealing is performed starting from a random spin configuration at each $H$,  i.e., gradually lowering of the temperature to the given $T$. 
For the $T$ dependences of the order parameters, the temperature is lowered from a high temperature, and for the thermal hysteresis properties, temperature is changed without the initialization of the spin configuration. 
At each $T$ and $H$, 400,000 MC steps are used for the measurement of the order parameters after 100,000 MC steps for the equilibration, and the average of each order parameter is taken over 12 and 48 independent simulations for the $H$ and $T$ dependences, respectively. 
To treat large systems and exclude the effect of edges, we tile replicas of the original system of $N=L^2=96\times96$ sites with periodic boundary conditions~\cite{Ruger, Horowitz}. We tile 2001$\times$2001 replicas. 
Throughout this paper, we use $k_{\rm B}=1$.

\section{Results and Discussion \label{results} }

\begin{figure}[thbp]
\begin{center}
\includegraphics[width = 8.0cm]{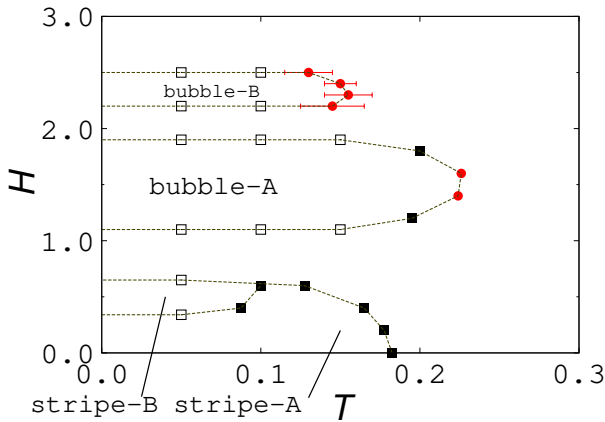}
\end{center} 
\caption{$H$-$T$ phase diagram of the system. The full circles show the first order transition points obtained by analyzing the energy histogram or thermal hysteresis property. The full squares are transition points estimated by the heat capacity analysis. The open squares are transition points estimated by the $H$ dependences of the order parameters. The error bars indicate the hysteresis widths.}
\label{PD1}
\end{figure}

\subsection{Phase diagram}
\label{phase_diag} 

\begin{figure}[hbp!]
\begin{center}
\begin{minipage}{0.99\hsize}
\includegraphics[width = 4.0cm]{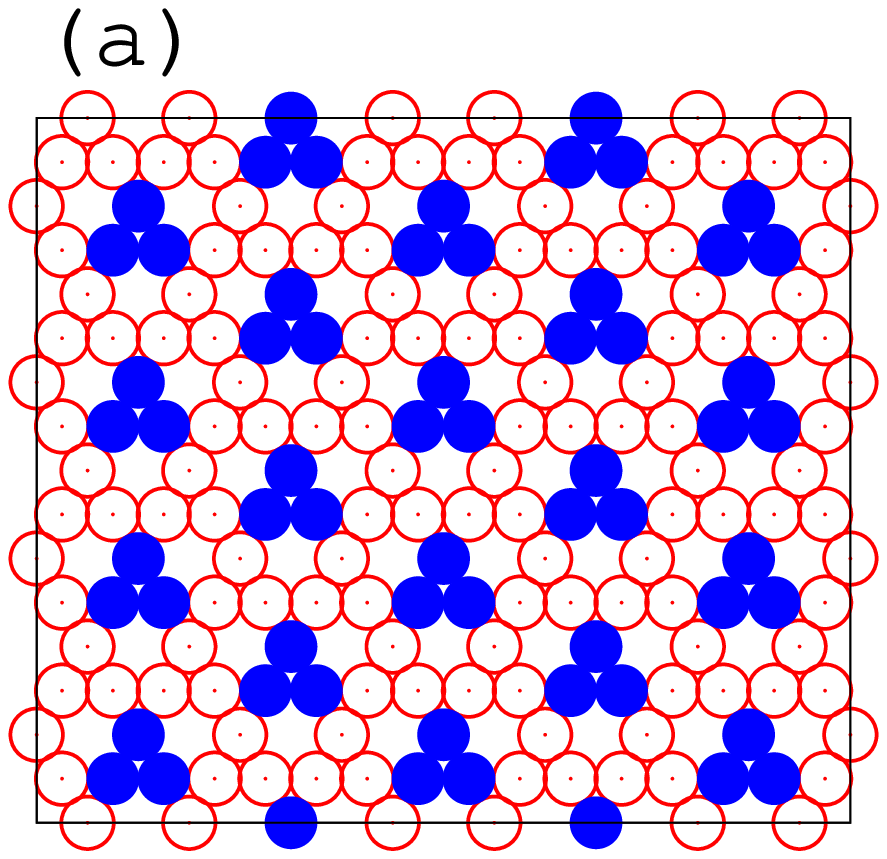} 
\includegraphics[width = 4.0cm]{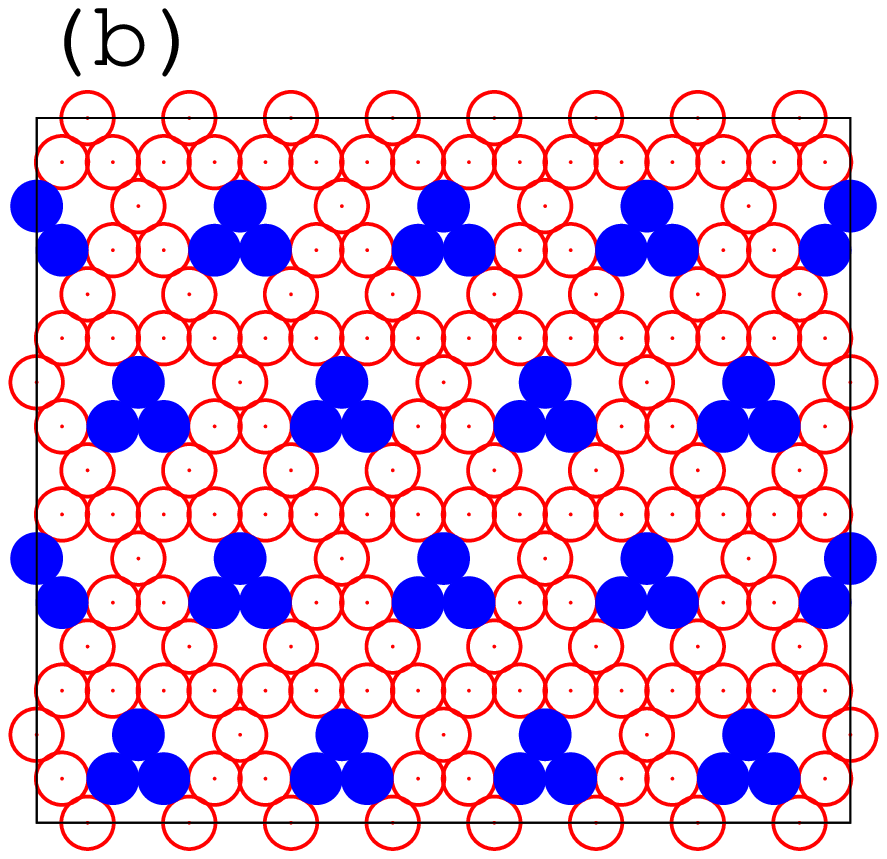} 
\end{minipage}
\end{center}
\caption{Snapshots of the magnetic structure at $T=0.05$ in the space filling configuration (a) in bubble phase A at $H=1.6$ and (b) in bubble phase B at $H=2.3$. 
16$\times$8$\sqrt{3}$-sized region is displayed. Blue full and red open circles denote down and up spins, respectively. }
\label{SS_b}
\end{figure}

First we give an overview of the structure of the phase diagram obtained in this study. The field--temperature ($H$-$T$) phase diagram is shown in Fig \ref{PD1}. 
We find two bubble phases at high fields. 
We call the bubble phases at lower and higher fields bubble phases A and B, respectively. 
Figures~\ref{SS_b} (a) and (b) show the magnetic structures in the space filling configuration in bubble phases A and B, respectively. 
In both bubble phases, three nearest-neighbor down spins form a domain, i.e., bubble, and bubbles form a regular triangular lattice, but the triangular array of bubbles is different and the distance $r$ between the nearest neighbor bubbles changes. 
In bubble phase A, the distance is $r=2\sqrt{3}$, while in bubble phase B, $r=4$. 
In bubble phase B the distance is larger and the bubble density is lower, which increases the magnetization and gains the Zeeman energy.

\begin{figure}[hbp!]
\begin{center}
\begin{minipage}{0.99\hsize}
\includegraphics[width = 4.0cm]{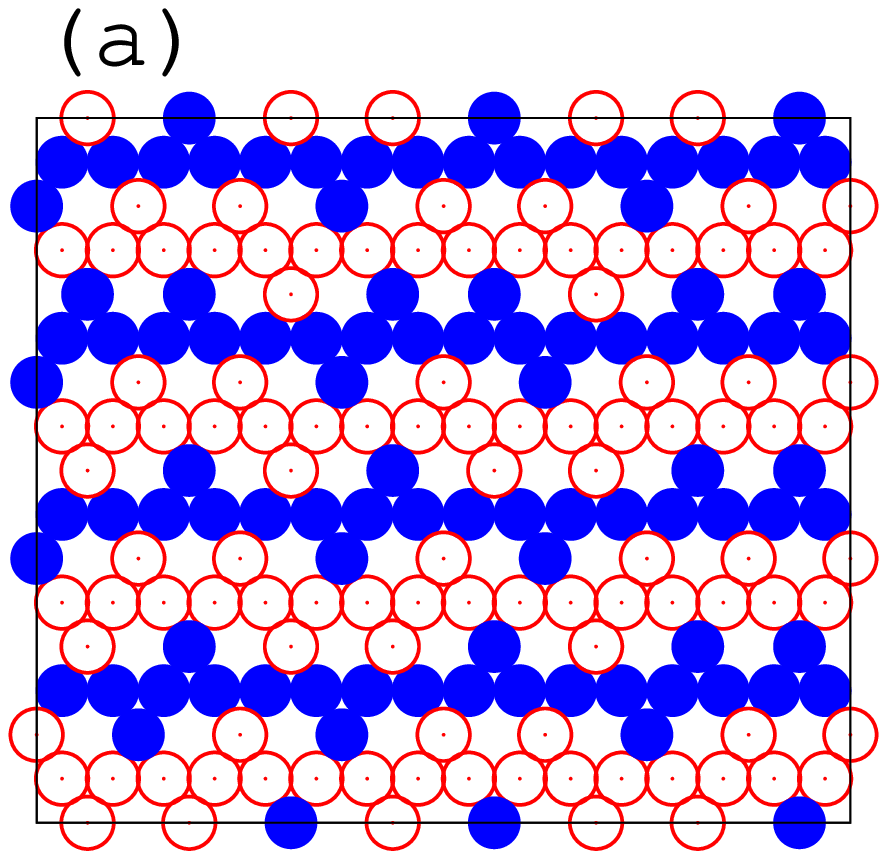}  
\includegraphics[width = 4.0cm]{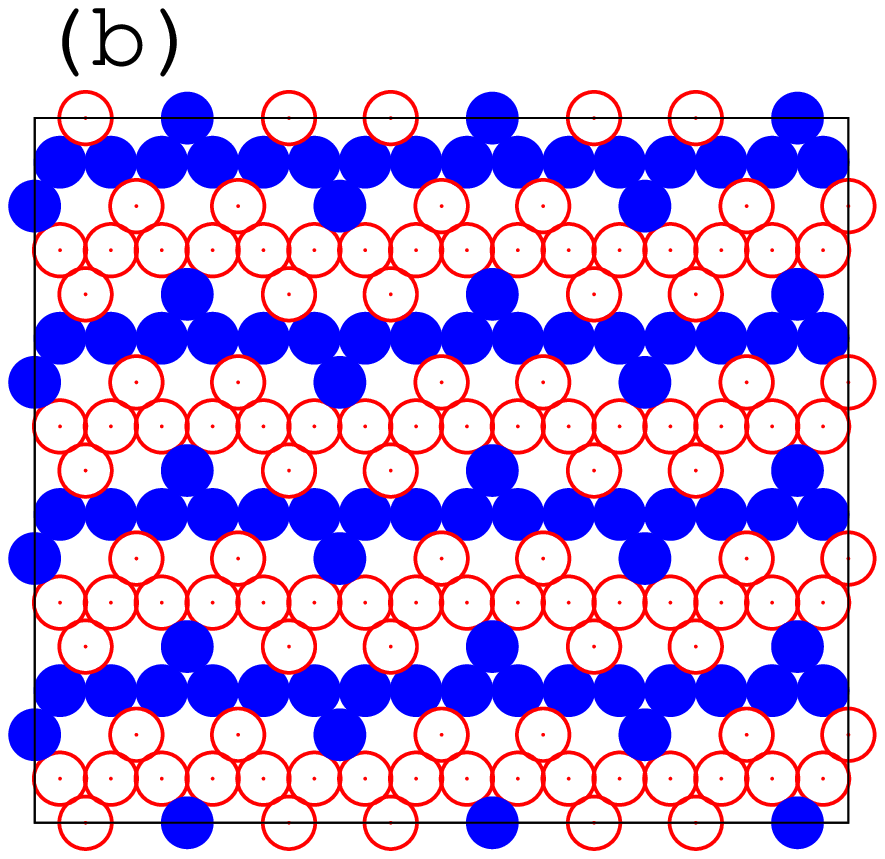} 
\includegraphics[width = 4.0cm]{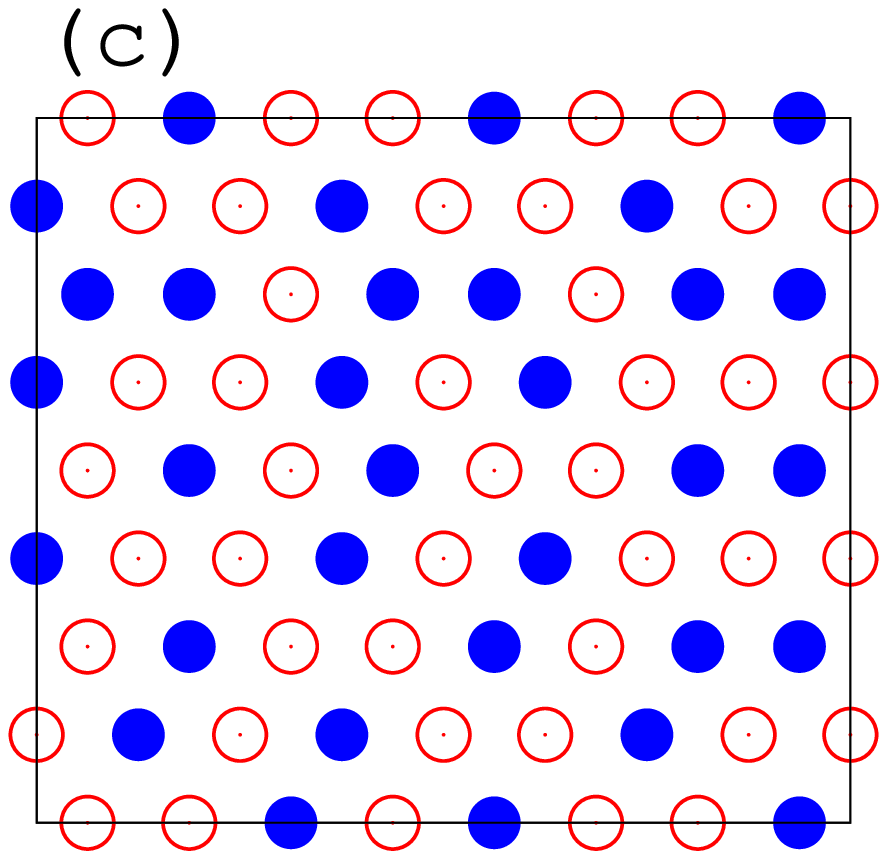}
\includegraphics[width = 4.0cm]{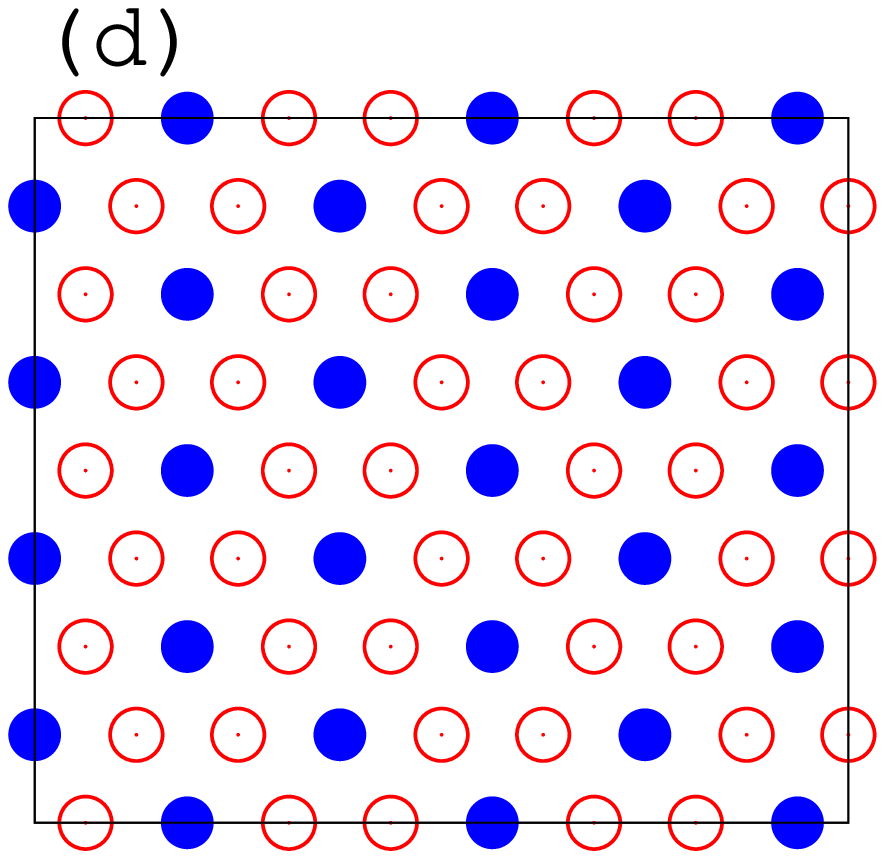}
\end{minipage}
\end{center}
\caption{Snapshots of the magnetic structure at $T=0.05$ in the space filling configuration (a) in stripe phase A at $H=0.2$ and (b) in stripe phase B at $H=0.6$. 
Snapshots of the magnetic structure of ``branch spins"(c) in stripe phase A and 
(d) in stripe phase B. 
16$\times$8$\sqrt{3}$-sized region is displayed.  Blue full and red open circles denote down and up spins, respectively. It should be noted that the branch spins in stripe phase B form a triangular lattice. }
\label{SS_s}
\end{figure}

We also find two phases at low fields, whose magnetic structure has a stripe frame of down spins with 1-stripe width (Figs.~\ref{SS_s} (a) and (b)). 
We call these phases at lower and higher fields stripe phases A and B, respectively. 
In the two phases, the location of the down spins neighboring the stripe frame (blue circles in Figs.~\ref{SS_s} (c) and (d)) is different. 
We call these neighboring spins ``branch spins".  
In the magnetic structure in stripe phase A illustrated in Fig.~\ref{SS_s} (a), the branch spins do not form a lattice and align randomly (Fig.\ref{SS_s} (c)). 
In stripe phase B, however, the branch spins form a triangular lattice (Fig.~\ref{SS_s} (d) ). 

\begin{figure}[htb]
\begin{center}
\begin{minipage}{0.99\hsize}
\includegraphics[width = 8.0cm]{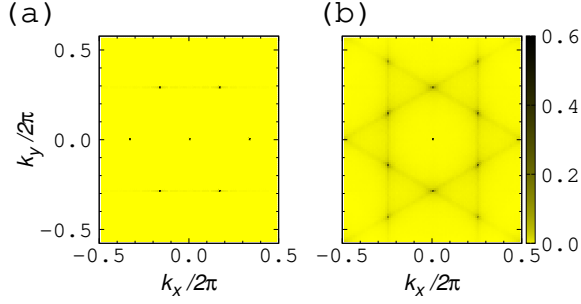} 
\end{minipage}
\end{center}
\caption{ $\bar{\varphi} _{\vector{k} } $ at $T=0.05$ at (a) $H=1.6$ and (b) $H=2.3$. }
\label{Frr1}
\end{figure}

\subsection{Bubble phases}
\label{bubble}
 
We introduce several types of order parameters to study the phase transitions in
the system. 
To investigate bubble phases A and B, we analyze the Fourier component of the spin configuration, 
\begin{equation}
\bar{\varphi} _{\vector{k} } =\Big\langle \Big | \frac{1}{N} \sum _{i}    \sigma_i   e^{ \vector{k} \cdot \vector{x}_i } \Big | \Big\rangle, 
\label{phi_gengeral}
\end{equation}
where $\langle \cdots \rangle$ denotes the thermal average. 
 
In Figs.~\ref{Frr1} (a) and (b), we illustrate $|\bar{\varphi} _{\vector{k} }|$ at $H=1.6$ and $H=2.3$, respectively, at $T=0.05$. 
The smallest reciprocal lattice vectors for the triangular lattice of bubbles in bubble phase A are  $\vector{k}_1 = (\frac{2 \pi}{3}, 0)$ and $\vector{k}'_1= (-\frac{\pi}{3}, \frac{\sqrt{3}\pi}{3})$, and those in bubble phase B are $\vector{k}_2 =(0,\frac{\pi}{\sqrt{3}})$ and $\vector{k}'_2 =(\frac{\pi}{2}, -\frac{\sqrt{3}\pi}{6})$. 
We find that linear combinations of $\vector{k}_1$ and $\vector{k}'_1$ and those 
of $\vector{k}_2$ and $\vector{k}'_2$ correspond to the high intensity parts (black spots) in Figs.\ref{Frr1} (a) and (b), respectively. 
Hence, we define the order parameters for bubble phase A,  
\begin{equation}
\varphi _{\rm 1} = \Big\langle   \Big | \frac{9}{ 4N } \sum _{j} \sigma _j   e^{i \frac{2 \pi }{3} x_j } \Big |  \Big\rangle,
\label{phi_1}
\end{equation}
and
\begin{equation}
\varphi'_{\rm 1} = \Big\langle \Big |  \frac{9}{ 4N } \sum_{j}  \sigma _j  e^{i (-\frac{\pi }{3} x_j+\frac{\sqrt{3} \pi }{3} y_j) }  \Big |  \Big\rangle,
\end{equation}
and for bubble phase B, 
\begin{equation}
\varphi _{\rm 2} =\Big\langle  \Big |  \frac{6}{ \sqrt{5} N } \sum _{j} \sigma _j e^{i \frac{\pi }{\sqrt{3} } y_j } \Big |  \Big\rangle,
\label{phi_2}
\end{equation}
and
\begin{equation}
\varphi'_{\rm 2} =\Big\langle  \Big |  \frac{6}{ \sqrt{5} N} \sum_{j} \sigma _j e^{i (\frac{\pi }{2} x_j-\frac{\sqrt{3} \pi }{6} y_j) } \Big |  \Big\rangle. 
\end{equation}
Here, the prefactors $\frac{9}{4}$ and $\frac{6}{\sqrt{5} }$ are normalization constants. 

We also study the magnetization of the system to characterizes the regions of the ordered phases, 
\begin{equation}
m_z=\Big\langle \frac{1}{N} \sum_i^N  \sigma_i \Big\rangle,
\end{equation}
and 
the sum of the Ising and dipolar interaction energies, 
\begin{equation}
E = \Big\langle \frac{1}{N} \Big(- \delta \sum _{\left< i,j \right> } \sigma _i \sigma _j  + \sum _{i < j  } \frac{ \sigma _i \sigma _j }{r_{ij} ^3} \Big)  \Big\rangle. 
\label{Energy}
\end{equation}

\begin{figure}[hbt]
\begin{center}
\begin{minipage}{0.99\hsize}
\includegraphics[width = 7.5cm]{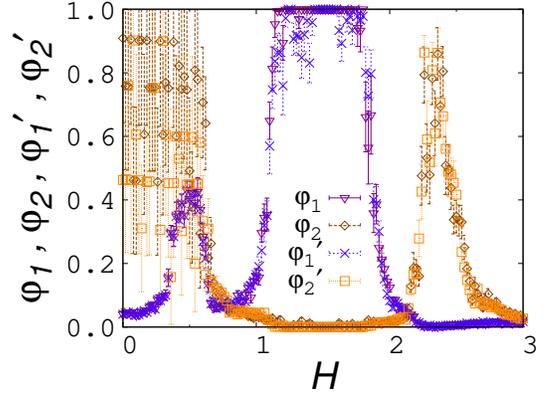} 
\end{minipage}
\end{center}
\caption{$H$ dependences of $\varphi _{\rm 1}$, $\varphi'_{\rm 1}$, $\varphi_{\rm 2}$, and $\varphi'_{\rm 2}$ at $T=0.05$.} 
\label{phi-H}
\end{figure}

\begin{figure}[hbp!]
\begin{center}
\includegraphics[width = 7.5cm]{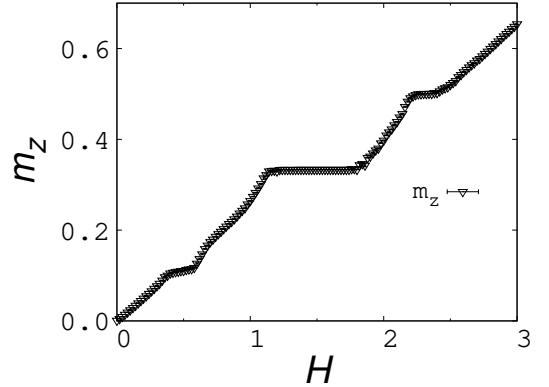}
\end{center}
\caption{$H$ dependence of $m_z$ at $T=0.05$. }
\label{mz-H}
\end{figure}

\begin{figure}[hbp!]
\begin{center}
\includegraphics[width = 7.5cm]{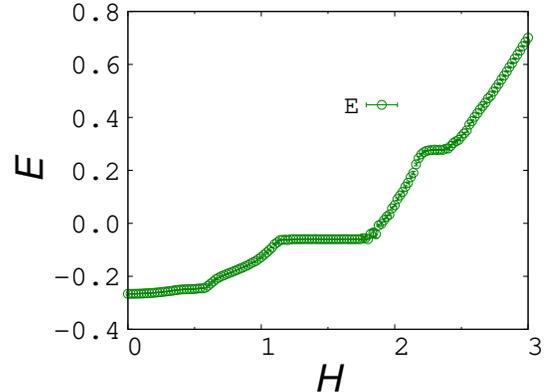}
\end{center}
\caption{$H$ dependence of $E$ at $T=0.05$.}
\label{E-H}
\end{figure}

We plot the $H$ dependences of $\varphi_{\rm 1}$, $\varphi'_{\rm 1}$, $\varphi_{\rm 2}$, and $\varphi'_{\rm 2}$ at $T=0.05$ in Fig.~\ref{phi-H}, and find that bubble phase A appears at $1.1 \lesssim H \lesssim 1.9$ and bubble phase B at $2.2 \lesssim H \lesssim 2.5$. 
In these field regions, plateaus of $m_z$ and $E$ appear in Figs.~\ref{mz-H} and \ref{E-H}, respectively, which indicates that the configurations of Figs.~\ref{SS_b} (a) and (b) are maintained. (Above $H=3$, $m_z$ monotonically increases and is saturated at $H \simeq 4.0$). 
  
We find that at $0.34 \lesssim H \lesssim 0.65$, $\varphi_{\rm 1}$ and $\varphi'_{\rm 1}$ appear but much less than 1 and a plateau of $m_z$ is accompanied, and at $H \lesssim 0.65$, $\varphi_{\rm 2}$ and $\varphi'_{\rm 2}$ appear with large fluctuation. 
We consider the reasons for these observations in the next subsection.

\begin{figure}[hbp!]
\begin{center}
\begin{minipage}{0.99\hsize}
\includegraphics[width = 7.5cm]{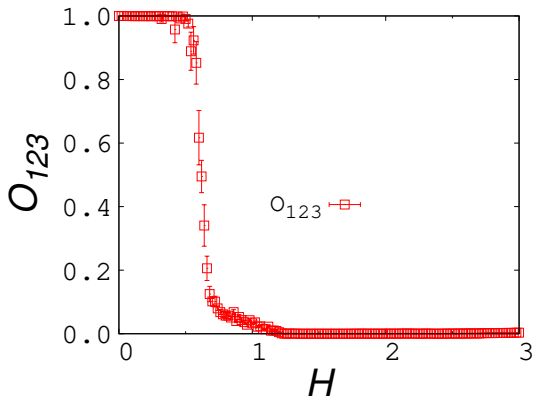}
\end{minipage}
\end{center}
\caption{$H$ dependence of $O_{\rm 123}$ at $T=0.05$. }
\label{H_O123}
\end{figure}

\begin{figure}[hbp!]
\begin{center}
\includegraphics[width = 7.5cm]{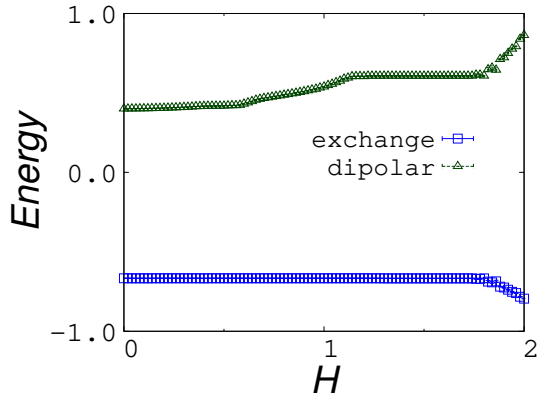}
\end{center}
\caption{$H$ dependences of the exchange and dipolar interaction energies at $T=0.05$.}
\label{Eex_Ed-H}
\end{figure}

\begin{figure}[hbp!]
\begin{center}
\includegraphics[width = 7.0cm]{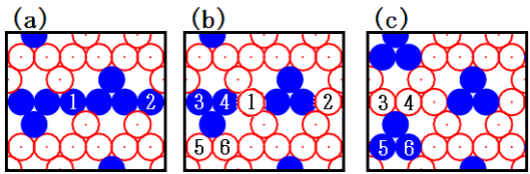}
\end{center}
\caption{Configurations with the same exchange interaction energy. 
The difference in the configuration between (a) and (b) is spins 1 and 2, and 
that between (b) and (c) is spins 3, 4, 5, and 6.}
\label{conf_same_ex}
\end{figure}

\begin{figure}[hbp!]
\begin{center}
\includegraphics[width = 4.0cm]{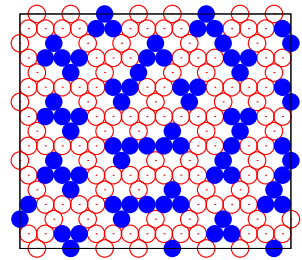}
\end{center}
\caption{Snapshot of the spin configuration at $H=0.9$ and $T=0.05$.}
\label{snap-H_0.9}
\end{figure}

\subsection{Stripe phases}
\label{stripe}

Stripe phases of Ising dipolar systems have been studied mainly on a square lattice. There the numbers of horizontal ($n_h$) and vertical ($n_v$) bonds between nearest-neighbor antialigned spins are calculated, and the order parameter is defined as the difference between $n_h$ and $n_v$, which detects $\pi/2$-rotational symmetry breaking~\cite{Booth,Cannas}. 

In the case of the kagome lattice, we should investigate $\pi/3$-rotational symmetry breaking to detect the stripe phases. 
There are three bond angles from the $x$-axis concerning nearest-neighbor antialigned spin pairs, i.e., $0, \frac{\pi}{3}$, and $\frac{2\pi}{3}$.  
We define the order parameter, 
\begin{equation}
O_{\rm 123} = 2 \frac{ \left| n_1 + n_2 \omega + n_3 \omega ^2 \right| }{ \left| n_1 + n_2 + n_3 \right| }. 
\label{O123}
\end{equation}
Here, $\omega = e^{\frac{2 \pi i}{3} } $, which satisfies $1+\omega+\omega ^2=0$,
and $n_1 , n_2$, and $n_3$ are the numbers of bonds of nearest-neighbor antialigned spins with the bond angles, $0, \frac{\pi}{3}$, and $\frac{2\pi}{3}$, respectively. The prefactor 2 is a normalization constant. 

We present the $H$ dependence of $O_{\rm 123}$ in Fig.~\ref{H_O123}. 
We find that at $H \lesssim 0.65$, $O_{\rm 123}$ reaches almost full saturation, and  a stripe phase or stripe phases are identified in this region. 
The period of the stripes is $2\sqrt{3}$ (Figs.~\ref{SS_s} (a) and (b)), and the horizontal stripes are detected by $\varphi_{\rm 2}$, and the diagonal stripes parallel to the direction of (1, $\sqrt{3}$) are detected by $\varphi'_{\rm 2}$. Therefore, finite values of $\varphi_{\rm 2}$ and $\varphi'_{\rm 2}$ with fluctuation in Fig.~\ref{phi-H} are ascribed to the formation of these stripes.  

We find in Figs.~\ref{mz-H} and \ref{E-H} a plateau-like region of $m_z$ and $E$ at $0.34 \lesssim H \lesssim 0.65$, at which $\varphi_{\rm 1}$ and $\varphi'_{\rm 1}$ have finite values ($<0.5$) in Fig.~\ref{phi-H}. This plateau-like region suggests that the structure of Fig.~\ref{SS_s} (b) is stable against the magnetic field. 
We notice that the branch spins in stripe phase B form a triangular lattice with 
lattice constant $2\sqrt{3}$, which is the same as that of the triangular bubble lattice in bubble phase A. 
The triangular lattice of the branch spins causes 
finite values of $\varphi_{\rm 1}$ and $\varphi'_{\rm 1}$, which is an evidence for 
the realization of stripe phase B at $0.34 \lesssim H \lesssim 0.65$. 

We find that the exchange interaction energy, i.e., the first term of $E$ is constant for $H \lesssim 1.8$ (Fig.~\ref{Eex_Ed-H}) including the regions of  stripe phases A and B, and bubble phase A. 
The constant exchange interaction energy is easily confirmed between the two stripe phases, and it is also confirmed between stripe phase B and bubble phase A by 
considering the transformations between (a) and (b) and between (b) and (c) in Fig.~\ref{conf_same_ex}. 
It is interesting to note that in the field region of the disordered phase between stripe phase B and bubble phase A, the exchange interaction energy is unchanged, although disordered spin configurations such as Fig.~\ref{snap-H_0.9} at $H=0.9$ and $T=0.05$ appear. 

In the two stripe phases, because the exchange energy associated with branch spins are zero and thus the exchange energy is originated only from the stripe part, the spin configuration of the branch spins is considered to be located on the triangular lattice with lattice constant $2\sqrt{3}$ only by the dipolar interaction and the magnetic field. 

\subsection{Phase transitions}
\label{PT}

\begin{figure}[hbp!]
\begin{center}
\begin{minipage}{0.99\hsize}
\includegraphics[width = 7.5cm]{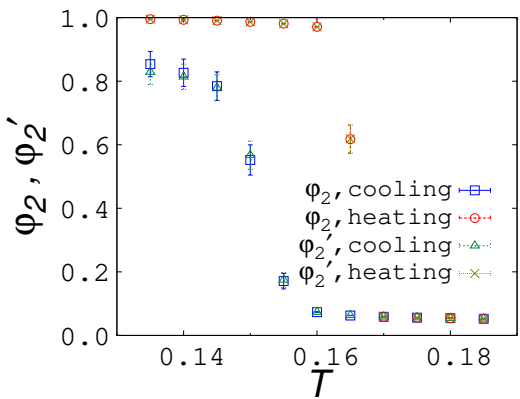} 
\end{minipage}
\end{center}
\caption{ $\varphi _2$ and $\varphi' _2$ in heating and cooling process at $H=2.3$. Hysteresis loops of $\varphi _2$ and $\varphi' _2$ are observed.} 
\label{Hysteresis_h2.3}
\end{figure}

We investigate the properties of phase transitions. 
In Fig.~\ref{Hysteresis_h2.3}, $\varphi _2$ and $\varphi' _2$ in a heating and cooling process at $H=2.3$ are shown. We find thermal hysteresis loops of $\varphi _2$ and $\varphi' _2$, which indicate the existence of metastable states between bubble phase B and the disordered phase. Therefore, the transition associated with these hysteresis loops is identified as a first order transition. 
We adopt the middle point (temperature) of the loops as the first order transition point, and plot this point with a red circle on the phase diagram of Fig.\ref{PD1}, in which the error bar coincides with the loop width. 
In the same manner, first-order transition points between bubble phase B and the disordered phase are plotted with red circles on the phase diagram.  

\begin{figure}[hbp!]
\begin{center}
\begin{minipage}{0.99\hsize}
\includegraphics[width = 7.5cm]{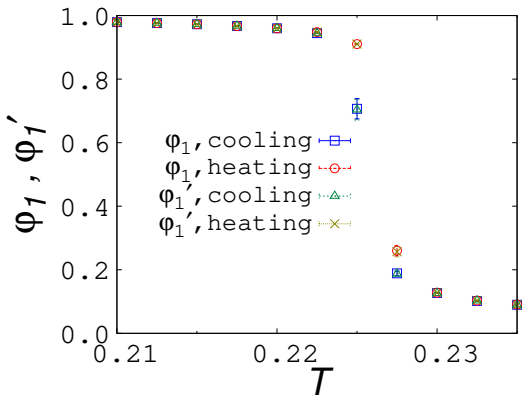} 
\end{minipage}
\end{center}
\caption{$\varphi _1$ and $\varphi' _1$ in heating and cooling process at $H=1.6$.} 
\label{Hysteresis_h1.6}
\end{figure}
\begin{figure}[hbp!]
\begin{center}
\begin{minipage}{0.99\hsize}
\includegraphics[width = 7.5cm]{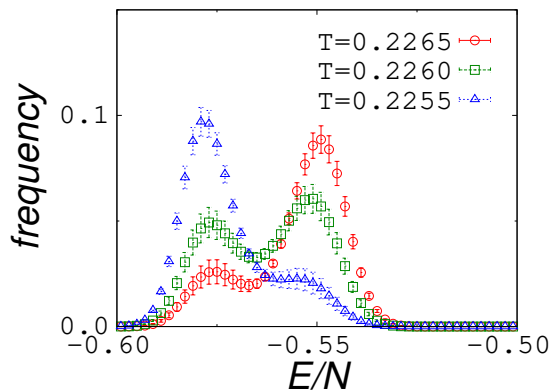} \\ 
\end{minipage}
\end{center}
\caption{Energy histograms at $H=1.6$ at $T=0.2265$, 0.226, and 0.2255.}
\label{Histogram}
\end{figure}

In Fig.~\ref{Hysteresis_h1.6}, $\varphi _1$ and $\varphi' _1$ in a heating and cooling process at $H=1.6$ are presented. We find a very small hysteresis loops but it is difficult to judge if the transition is of first order. Then, we perform an energy histogram analysis.  If the phase transition
is of first order, the histogram should have two peaks
around the transition temperature. 
We find double peaks around $T=0.226$ in Fig.\ref{Histogram} and judge this point to be a first-order transition point between bubble phase A and the disordered phase. In the same manner, the first-order transition point at $H=1.4$ is determined. 
These first-order transitions points are plotted by red circles between bubble phase A and the disordered phase on the phase diagram. 

\begin{figure}[hbp!]
\begin{center}
\begin{minipage}{0.99\hsize}
\includegraphics[width = 7.5cm]{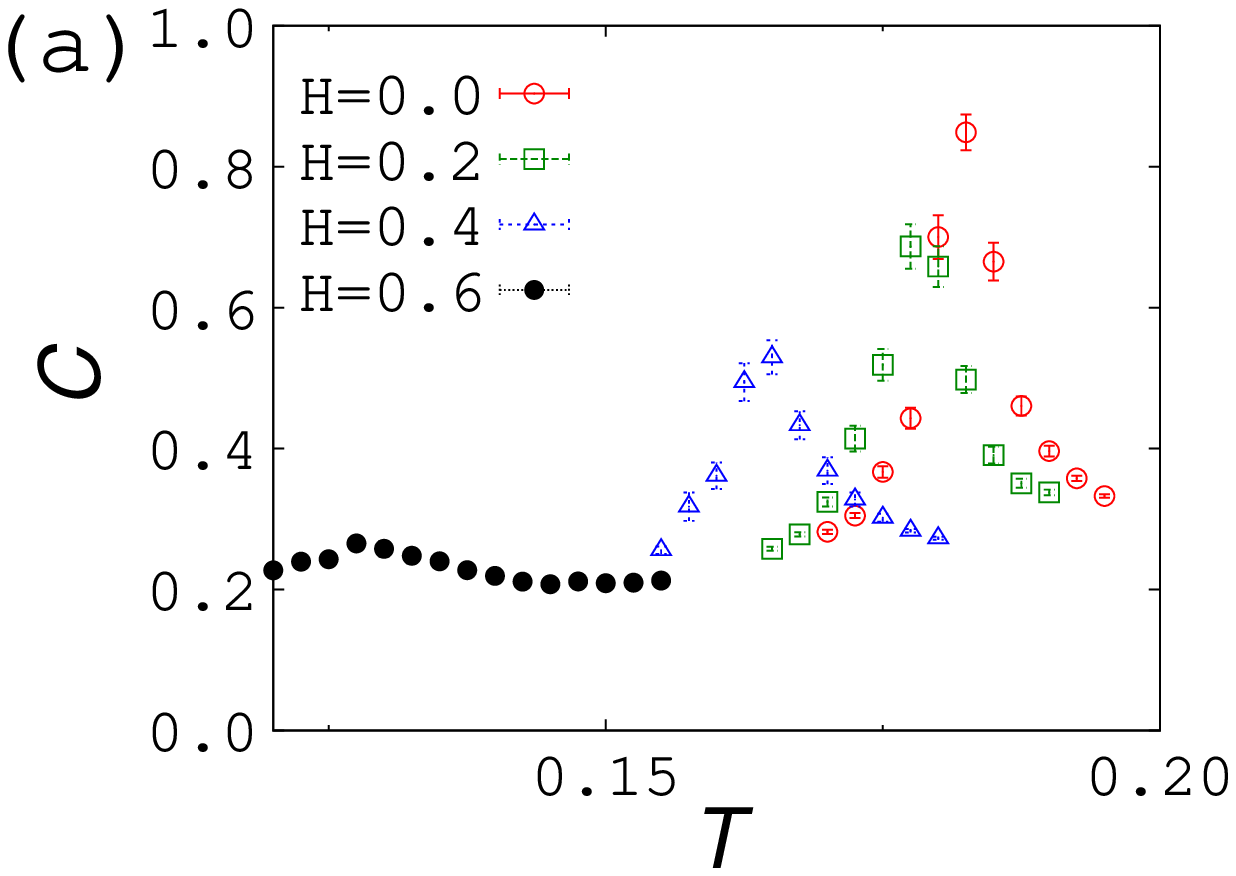} \\ 
\includegraphics[width = 7.5cm]{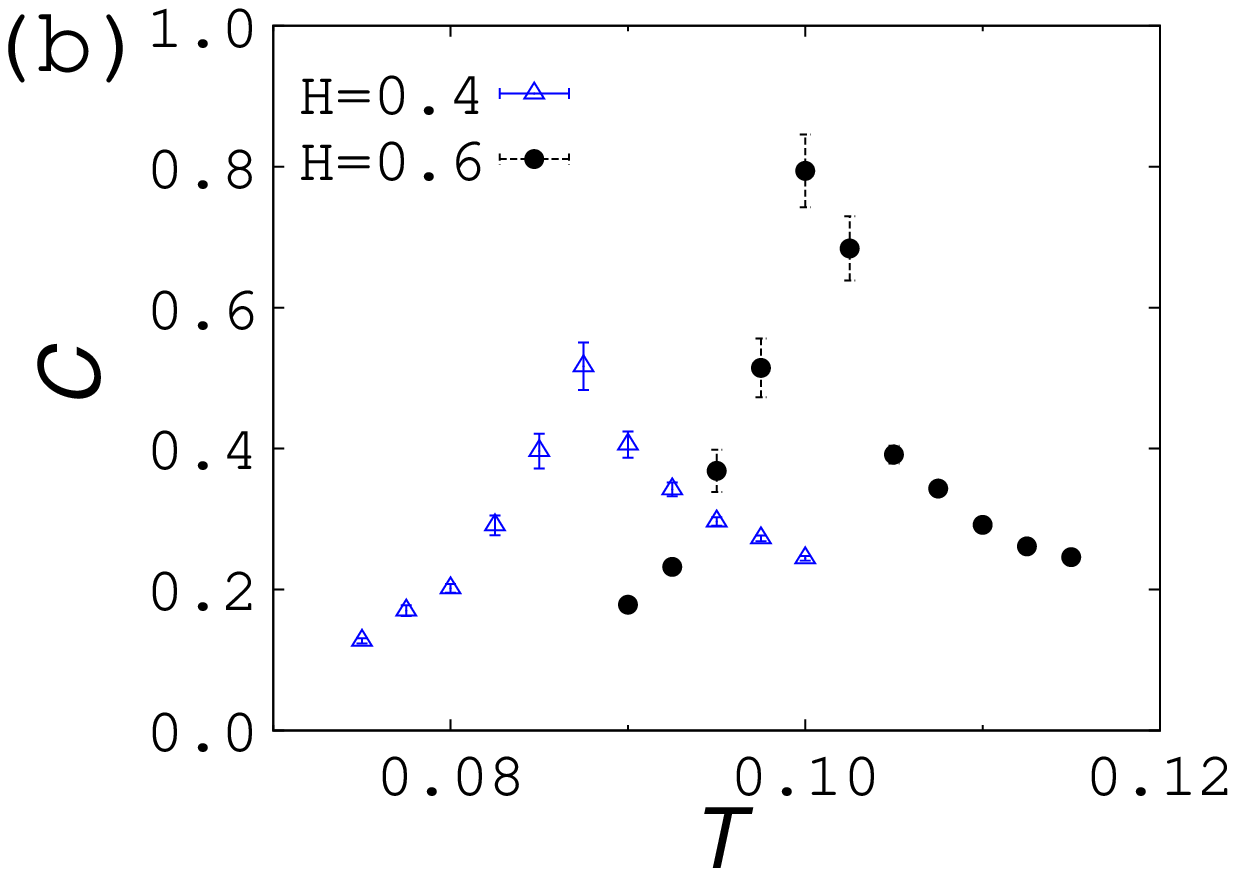}
\end{minipage}
\end{center}
\caption{Temperature dependence of heat capacity at (a) $H=0.0$, 0.2, 0.4, and 0.6 for $0.12 <T<0.2$,  and at (b) $H=0.4$ and $H=0.6$ for $0.07 <T<0.12$.} 
\label{heat_cap}
\end{figure}

We find points around phase boundaries, in which both of hysteresis loops and double peaks of the energy histogram have not been observed within the accuracy of the present work, and study the heat capacity per spin for several such points, 
\begin{equation}
C =\frac{1}{N} \frac{\langle E_t^2 \rangle-\langle E_t \rangle^2}{k_{\rm B} T^2 }, 
\label{cap}
\end{equation}
where $E_t$ is the total energy of the system. 
In Figs.\ref{heat_cap} (a) and (b), we give the temperature dependence of the heat capacity at $H=0.0$, 0.2, 0.4 and 0.6 for $0.12 <T<0.2$, and at $H=0.4$ and 0.6 for $0.07 <T<0.12$, respectively. 
The peaks of $C$ in Fig.\ref{heat_cap} (a) show the transition points between 
stripe phase A and the disordered phase, while the peaks of $C$ at $H=0.4$ and 0.6 in Fig.\ref{heat_cap} (b) give the transition points between stripe phases A and B. These points are plotted with full squares on the phase diagram.

We also analyze the temperature dependence of $C$ at $H=1.2$ and 1.8 between bubble phase A and the disordered phase and find peaks. 
These points are also plotted with full squares on the phase diagram. 
Considering the first-order transition points (red circles) at $H=1.4$ and 1.6, these transitions are of weak first order.  

\begin{figure}[hbp!]
\begin{center}
\begin{minipage}{0.99\hsize}
\includegraphics[width = 7.5cm]{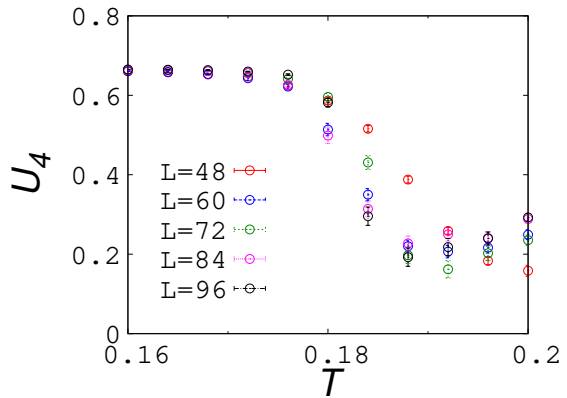} 
\end{minipage}
\end{center}
\caption{Temperature dependence of Binder cumulant at $H=0$.} 
\label{Binder_plot}
\end{figure}

In our previous study on a square lattice~\cite{Komatsu1}, the transition between the stripe and disordered phases at low $H$ was of second order. Here, to investigate the possibility of a second order transition, we calculate the Binder cumulant, $U_4$, of $O_{123}$ at low fields, 
\begin{equation}
U_4=1-\frac{\langle O_{123}^4 \rangle}{3\langle O_{123}^2 \rangle}. 
\end{equation}
In Fig.~\ref{Binder_plot}, the Binder cumulants at $H=0$ for different system sizes are plotted. 
We find no crossing of these cumulants and observe the same tendency at close squares at $H=0.2$ and 0.4 on the phase diagram, and we do not judge that the phase transition between stripe phase A and the disordered phase are of second order.
 
We give the phase boundaries at $T=0.05$ with open squares in the phase diagram 
judging from the 
$H$ dependences of the order parameters (Figs.~\ref{phi-H}, \ref{mz-H}, and \ref{H_O123}).
In the same manner, the other open squares are plotted using the $H$ dependences of the order parameters at $T=0.1$ and $T=0.15$. We find that a disordered phase exists between stripe phase B (or A) and bubble phase A, between bubble phases A and B, and above bubble phase B, and the phase boundaries at low temperatures extend to higher temperatures in $T$ direction. 

\section{Summary \label{summary} }

We investigated the two-dimensional dipolar Ising ferromagnet on a kagome lattice with a specific ratio between the exchange and dipolar constants, $\delta = 1$. 
We calculated the order parameters for bubbles and stripes and other thermal quantities using the stochastic cutoff $O(N)$ Monte Carlo method, and analyzed the field-temperature phase diagram. 

We found two stripe phases at low fields, where the stripe frame has a 1-stripe width. In the lower-field stripe phase (stripe phase A), the branch spins, defined as reversed spins neighboring the stripe frame, align randomly, while in the higher-field stripe phase (stripe phase B), the branch spins form a triangular lattice, which shows a magnetization plateau. In stripe phase A, the magnetization changes with increasing the field, but the exchange interaction energy is constant in both stripe phases A and B, where the dipolar interaction is essential in the formation of the magnetic structure of the branch spins.    

We also found two bubble phases at high fields. 
In both bubble phases, three nearest-neighbor down spins form a triangular domain, i.e., bubble, and bubbles form a triangular lattice, but the triangular array of bubbles and distance between bubbles vary. Interestingly, the exchange energy is constant not only in the two stripe phases but also in the lower-field bubble phase (bubble phase A) and intermediate disordered phase. 

We determined the phase boundaries and showed several properties of the phase transitions.  So far, a first-order transition was suggested between stripe and bubble phases. However, a specific lattice structure was assumed in the effective free energy approaches and bubble phases were studied without defining an order parameter for a lattice structure.  
In this paper we defined bubble phase as a phase with a lattice structure formed by magnetic domains. Consequently, we discovered that there exists a disordered phase between stripe A (B) and bubble phase A and between the two bubble phases, and the transitions between the two bubble phases and disordered phase are of first order at high temperatures. 

\begin{acknowledgments}
The present study was supported by Grants-in-Aid for Scientific Research C (No. 18K03444 and No. 20K03809) from the Ministry of Education, Culture, Sports, Science and Technology (MEXT) of Japan, and the Elements Strategy Initiative Center for Magnetic Materials (ESICMM) (Grant No. 12016013) funded by MEXT. 
The calculations were partially performed using the Numerical Materials Simulator (supercomputer) at the National Institute for Materials Science.  
\end{acknowledgments}

\end{document}